# London Quantum-Secured Metro Network

Andrew Lord[1], Robert Woodward[2], Shinya Murai[3], Hideaki Sato[3], James Dynes[2], Paul Wright[1], Catherine White[1], Russell Davey[1], Mark Wilkinson[1], Piers Clinton-Tarestad[4], Ian Hawkins[1], Kristopher Farrington[1], Andrew Shields[2]

[1] BT Adastral Park, Ipswich, UK, [2] Toshiba Europe Ltd., Cambridge, UK, [3] Toshiba Digital Solutions Ltd., Tokyo, Japan, [4] EY, London, UK
andrew.lord@bt.com

**Abstract:** We describe a London Quantum-Secured Metro Network using Quantum Key Distribution between three London nodes together with customer access tails. The commercially-ready solution is fully integrated into the BT network and on-boarded its first customer. ©2022 The Author(s)

## 1. Introduction

Quantum Key Distribution (QKD) has made sufficient progress to now be implemented in a fully commercial setting. There have been numerous examples of point-to-point commercial solutions and the next step is to start to commercialize QKD networks, in which the core key distribution is shared over multiple end customers, for a cost-effective and efficient solution. In this paper we describe a commercially-ready QKD metro network built in London, complete with customer access tails and an aggregated central metro node, able to support multiple customers. The solution includes a full Key Management System; encrypted classical Ethernet data on the same fibre and a Data Communications Network (DCN) for full remote monitoring at BT's Network Operations Centre (NOC)

## 2. Network overview

*2.1 Overall design*

Fig.1 shows the node locations for the London Quantum Secured Metro Network (L-QSMN), comprising three metro nodes, located in three of the main BT exchanges in the London metro area. These nodes are all interconnected with both QKD systems and WDM, co-existing on the same fibre pair. Customers within an up to 30km radius of one of these nodes will be able to join the trial via a dedicated access QKD connection, also with co-existing, encrypted ethernet.

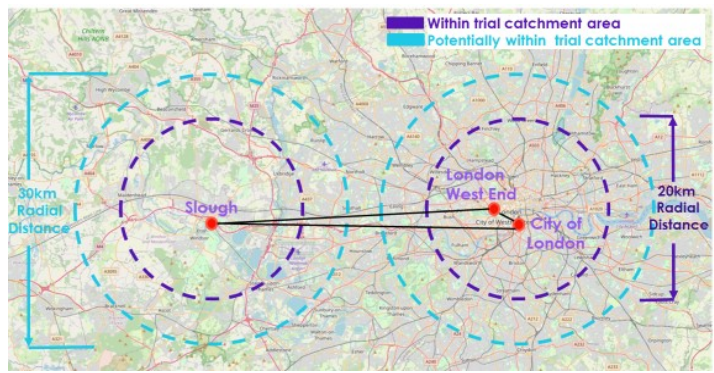

Fig.1 Map of London Quantum Metro Network

*2.2 Architecture*

Fig.2 describes the top level architecture, with several key features:
(i) QKD keys are established separately between (1) Customer Site A – Core Site A, (2) Core Site A – Core Site B and (3) Core Site B – Customer Site B. The keys produced in (2) and (3) are used to carry the keys produced in (1) to the far end of the circuit at Customer Site B. This is effected via information-theoretically-secure (one-time-pad based) symmetric key encapsulation, using one-time-use QKD keys from (2) and (3) to encapsulate the symmetric keys from (1), which are used to derive the final key which encrypts the data.

(ii) A Key Management Server (KMS) is placed at each node in the network and its role is to carry out the steps described in (i) and perform overall key management for the entire network.
(iii) Unencrypted customer data is encrypted on entry to the WDM system, using both internally generated keys from the WDM equipment and external keys fed from the KMS. A final encryption key is derived at the encryptor (which is a 10Gbps transmission card with AES-256 encryption) using a key derivation function which combines the entropy distributed by the KMS layer (using QKD link keys), with key material established between the endpoints using DH (Diffie-Hellmann) key exchange.
(iv) The data is optically terminated at the Core Sites because the access tail uses a different optical transmission system to the core. However the data payload is not decrypted until arriving at Customer Site B.

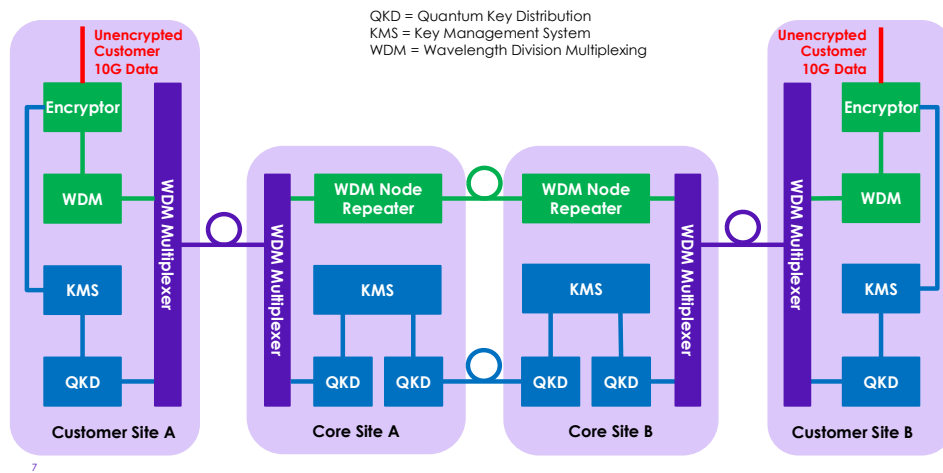

Fig.2 QKD network high level architecture

*2.3 Customer access tails*
A dark spectrum product [1] is used for the commercial access tails. Here, 8 wavelengths are provided to the customer via a wavelength multiplexer. One of these wavelengths is used for the QKD connection, whilst some of the others are used for the Ethernet data channel together with the QKD classical communication and synchronization channels.

*2.4 Metro network design*
In the metro network, core fibres are used exclusively to distribute the QKD quantum, synchronization and discussion channels. Other fibres are used to distribute the data channels (the OTN circuits). This prevents any need to restrict the range of the QKD based on co-propagation of large amounts of classical data [2], and maximises the possible distance between core nodes. For each core node, the site of the trusted node handover is a secure location within a BT exchange.

**3. Major technologies**

*3.1 QKD technology*
A combination of QKD systems are used in the network. For the core network, both Multiplexed Unidirectional (MU) and Long Distance (LD) QKD systems are used. The MU version features a quantum channel at ~1310nm which allows the propagation of high launch powers of data traffic in the same fibre. The system can be configured to use just one fibre to save fibre leasing costs, but here we adopted the co-propagating configuration which uses two fibres – one for all forward directed traffic and a second for all backward directed traffic. The LD version features a quantum channel aligned to the Dense Wavelength Division Multiplexing DWDM channel 34 (1550.12nm) and passes the quantum signals over a dark fibre. The range of the Long Distance system is typically 120km of standard single mode fibre.  For realising QKD in the access networks, the QKD system's quantum and classical data wavelengths have to be aligned to the Openreach specification for access equipment, Optical Spectrum Access (OSA) Filter Connect [3]. All wavelengths were then (de)muxed using external multiplexors conforming OSA criteria.  All QKD systems described above utilise an efficient version of the standard BB84 one-way protocol with decoy states [4]. Encoding and decoding of the keys is accomplished through the use of phase modulation based on built-in asymmetric Mach-Zehnder interferometers. Avalanche photodiodes operated in Geiger mode in the QKD receiver comprise the single photon detectors and single photon avalanche extraction uses the self-differencing technique [5]. Where appropriate the systems utilise a combination of both spectral and time-domain filtering at the QKD receiver to isolate the quantum channel from co-propagating / Raman-scattered optical power [6]. Continuous operation of the QKD systems is achieved through feedback algorithms which constantly optimise the polarisation, phase and temporal delay of the exchanged photons. Finite effects are included as standard with a resulting key failure probability of $10^{-10}$. Finally both the QKD transmitter and receiver fit into a standard 19 inch rack and take up three rack units (3U) of rack space.

*3.2 Key Management Layer*

The key management layer is a fully integrated solution providing a high speed network key delivery and management system for both the core and access networks. The key management layer collects keys from the QKD systems and uses these to one-time-pad encrypt global keys [7]. These global keys can then be routed around the network as required. The keys can be accessed by customer applications using a REpresentational State Transfer (REST-based) application interface (API). This API has been standardised by the European Telecommunications Standards Institute (ETSI) known as ETSI GS QKD 014 [8]. QKD key API standardisation is important as it allows application vendors to seamlessly integrate QKD key request interfaces into their technology.

**3. Network Integration**

The overall solution has been integrated into BT's DCN. Performance, faults and alarms are sent automatically to the Network Operations Centre, as is essential in order to provide carrier-grade availability and operational support. The elements in the core sites are directly connected to a dedicated management (DCN) network allowing them to be remotely managed via secure protocols and for elements to send Simple Network Management Protocol (SNMPv3) traps for status and alarm reporting. Network elements at customer sites use in-band management to relay configuration and alarms through to centralized servers via their connected nodes at the core sites.

**4. Performance and Use-cases**

The network has been fully commissioned, and has been operating stably for several months.

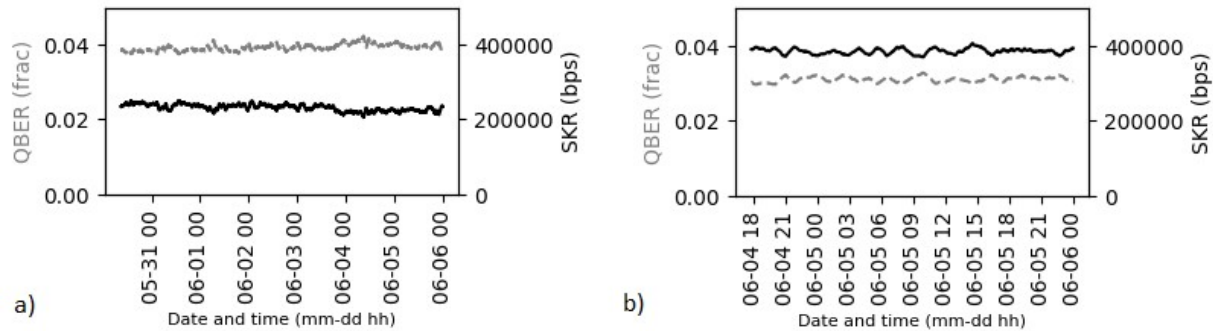

Fig.3 Representative QKD link performance on (a) an access link (length 12.0km) and (b) core link (length 46.7km)

The maximum key rate between the two customer sites is far more than sufficient to deliver the current requirement for one end-to-end 512 bit key refresh per minute at the encryptor, and therefore this system could provide independent keys for many applications. The first implemented use case is 10Gbit/s Ethernet encryption between customer sites and this has been installed for EY to allow them to send sensitive data, including videoconferencing, between their sites. Other use cases will be developed over the project lifespan.

**4. Conclusions and outlook**

A commercial QKD-based secure network around London has been integrated into BT's operational network. An encrypted Ethernet use case has been installed for the first customer on the trial. A novel Key Management System keeps track of keys and sends keys generated at one customer site to the receiving customer site. The next steps will focus on onboarding other customers, and investigation of broader use cases. Following this, it is expected that satellite-based QKD and long-haul terrestrial QKD will enable longer distance QKD links, which in turn will lead to interconnection between metro QKD islands.